# Alternation and interchange of e/4 and e/2 period interference oscillations as evidence for filling factor 5/2 non-Abelian quasiparticles

R.L. Willett*, L.N. Pfeiffer, K.W. West
Bell Laboratories, Alcatel-Lucent
600 Mountain Avenue
Murray Hill, New Jersey, 07974 USA
Correspondence to: rlw@alcatel-lucent.com

ABSTRACT:
   It is a theoretical conjecture that 5/2 fractional quantum Hall state charge e/4 excitations may obey non-Abelian statistics. In edge state interference these purported non-Abelian quasiparticles should display period e/4 Aharonov-Bohm oscillations if the interfering quasiparticle encircles an even number of localized e/4 charges, but suppression of oscillations if an odd number is encircled. To test this hypothesis, here we perform swept area interference measurements at 5/2. We observe an alternating pattern of e/4 and e/2 period oscillations in resistance. This aperiodic alternation is consistent with proposed non-Abelian properties: the e/4 oscillations occur for encircling an even number of localized quasiparticles, e/2 oscillations are expressed when encircling an odd number. Aperiodic alternation corresponds to the expected area sweep sampling the localized quasiparticles. Importantly, adding localized quasiparticles to the encircled area by changing magnetic field induces interchange of the e/4 and e/2 oscillation periods, specifically consistent with non-Abelian e/4 quasiparticles.

   The enigmatic fractional quantum Hall effect (FQHE) state at 5/2 filling factor [1-7] may possess excitations that obey non-Abelian statistics [2, 6-8]. If this is indeed true, these excitations may then be employable in performing topological quantum computation operations [9].
   It has long been postulated that the quasiparticle excitations in the fractional quantum Hall effect have non-trivial braiding statistics [10], with these quasiparticle charges and statistics potentially measurable in interference experiments [11-15]. With respect to the 5/2 state, theory has been proposed [16-20] that an interference experiment may be able to discern it's statistics question; if 5/2 quasiparticles can be made to encircle flux quanta and localized quasiparticles in an interference device, a distinctive pattern should result if the quasiparticles are non-Abelian. For an even number of encircled, localized particles, e/4 period Aharonov-Bohm (A-B) oscillations should result. However, if an odd number of quasiparticles are encircled these oscillations will be suppressed. Experimentally, interferometry of FQHE edge states, and in particular 5/2, has been accomplished recently [21]. The finding of interference periods consistent with e/4 charge in that study corroborates reports of the determination of 5/2 excitation charge [22-23], but charge determination does not define the statistics of the excitations. The question of the 5/2 quasiparticle statistics warrants an interference measurement that demonstrates the consequence of enclosing even and odd numbers of quasiparticles.

We report here interference measurements at filling factor ν=5/2 that address this hypothesis. An interferometer with an adjustable side gate is used that changes the encircled area, changing the number of enclosed flux quanta, and for large enough excursion changes the number of encircled, localized quasiparticles. We observe for large gate excursions an aperiodic alternation of interference patterns; a distinct e/4 period oscillation alternates with a distinct e/2 period oscillation. This alternation can be effected by addition of magnetic flux, and its consequent quasiparticle number change, whereby the e/4 and e/2 periods interchange. These results are specifically consistent with non-Abelian e/4 quasiparticle properties in the response to both B-field application and gate excursion. The e/4 oscillations represent the prescribed encircling of an even number of localized quasiparticles. The alternate e/2 oscillations occur when the encircled area contains an odd number of quasiparticles as the e/4 oscillations are suppressed. The alternation pattern is exposed for sufficiently large area sweeps; the aperiodic pattern corresponds to the expected random spatial distribution of localized quasiparticles within the swept area. Importantly, these specific aperiodic patterns can be induced to interchange the e/4 and e/2 periods upon addition of magnetic field corresponding to addition of a single localized e/4 quasiparticle. The observation of both the alternation and interchange of e/4 and e/2 periods specifically supports the conclusion of non-Abelian statistics for the e/4 quasiparticle.

High mobility, high density heterostructures of fixed density n access filling factors ν=nh/eB through swept magnetic field B. See methods section in Supplementary Information for sample and data analysis details. The interferometer, shown schematically in Figure 1a, is composed of two quantum point contacts (qpcs) and a central channel region. Each component is independently controllable [21, 24]. The top gate structure is displayed in the electron-micrograph in Figure 1b inset. Transport is measured through the devices with $R_L$ [25] as defined in Figure 1; typical bulk and $R_L$ magneto-transport are displayed in Supplementary Information Figure S1. The qpcs' voltages are adjusted beyond full depletion to promote preservation of the 5/2 state in the interferometer, and the side gate voltage Vs is independently varied to change the enclosed area.

A-B oscillations will occur when the side gate voltage is varied, with the period P~ΔVs~ΔA~(h/e*)/B, Vs the side gate voltage, A the enclosed interferometer area, e* the interfering charge value, and B the magnetic field. Side gate voltage sweeps will change the enclosed number of flux quanta resulting in periods defined by this dependence on e* and B. As noted in prior work the period measured at ν=2 with e*=1 and at ν=5/3 or 7/3 with e*=e/3 and their respective B-field values can be used to derive the proportionality to establish the expected period for 5/2. This interference and successful determination of oscillation periods for multiple FQHE states has been demonstrated previously [21]. Oscillation periods at 5/2 consistent with e*=e/4 and e/2 were demonstrated in these studies over small Vs excursions.

The experimental focus in this study is on a) preparation of the sample to optimize observation of the e/4 and e/2 oscillations, b) markedly expanding the range of the Vs sweeps compared to previous studies [21] to determine the relative prevalence of the e/4 and e/2 oscillations, and c) examining the B-field dependence of the e/4 and e/2 oscillations near 5/2. As shown below, expanding the range of Vs sweep reveals e/4 and e/2 oscillation periods that alternate aperiodically. This alternation can then be made to interchange e/4 and e/2 periods with change in B-field corresponding to the addition of a single localized quasiparticle.

In Figure 1 interference patterns are studied at a series of filling factors. In panel c, transmission resistance $\Delta R_L$ is plotted against side gate voltage change at filling factor $\nu=2$. $\Delta R_L$ is the measured resistance with the gross background subtracted. As observed previously [21], A-B oscillatory features are present in $\Delta R_L$ with the dominant periodicity marked by vertical lines. The period of these vertical lines is determined by the spectral peak of a fast Fourier transform (FFT) performed on the $\Delta R_L$ data of Figure 1c; the inset displays the FFT result. This period of A-B oscillations is then used with the relationship $P \sim \Delta V_s \sim \Delta A = (h/e^*)/B$; assigning $e^*=e$ for $\nu=2$, the period for other filling factors can be derived using their expected charges. Such a derivation is made for the filling factor $\nu=7/3$, necessarily using both $e^*=e/3$ and the B-field values of the respective measurements near $\nu=2$ and $\nu=7/3$. $\Delta R_L$ is then measured versus change in side gate voltage near filling factor 7/3, with results displayed in Figure 1d. The vertical lines of panel 1d reflect the derived periodicity, and the A-B oscillations marked e/3 in the data at 7/3 are in agreement with that periodicity. This period is further corroborated by the spectral peak of an FFT performed on the Figure 1d $\Delta R_L$ data: see inset. The $\nu=2$ and 7/3 FFT spectra periods are plotted in panel 1b, demonstrating the proportionality of the product of the period and the B-field position of the measurement versus the inverse charge. A property of the interference oscillations observed with these interferometers as noted in previous studies [21] is that they occur in series with phase disruptions. To delineate the series of phase coherent oscillations, in all $\Delta R_L$ data the charge is marked for each coherent section: for example, Figure 1d has two coherent sections marked e/3.

The potential for interferometer resistance oscillations to be due to Coulomb blockade (C-B) rather than A-B interference has been studied [26], particularly for small interferometers similar to those used here. This possibility is contradicted here by the period dependence of Figure 1d and also in interference results at a series of integer quantum Hall states from 2 to 8, with data demonstrated in Supplementary Figure S2. At integer $\nu$, contrary to the B independent period demonstrated in C-B dominated interferometers [26], the oscillation periods of this study demonstrate linear increase with 1/B in side gate sweep measurements as specifically expected for A-B oscillations with a single charge. In addition, for swept B-field measurements, oscillations with B-field period independent of $\nu$ are observed near these integer filling factors, again consistent with A-B oscillations and contrary to the properties of Coulomb blockade. The presence of A-B oscillations and not Coulomb blockade in this study may be attributed to the open geometry of the quantum point contacts of the devices employed here.

Focus now centers on resistance $\Delta R_L$ data taken near filling factor 5/2, and as shown previously [21] A-B oscillations are observed that are consistent with e/4 and e/2 charges. A typical interferometer resistance trace is shown in panel 1e, where two predominant periods present serially as the side gate is swept. From Vs change of 0 to 110mV a larger period oscillation is present, and FFT of this range is shown in panel 1f. This period, as determined from the FFT peak, is consistent with A-B oscillations of charge e/4: the product of period and B-field position are plotted in panel 1b and fall in line with the $\nu=2$ and 7/3 periods. The $\Delta R_L$ data section of smaller period (110 to 200mV) and the peak of the FFT spectrum taken here (panel 1g) display a period consistent with e/2, also plotted in 1b.

Figure 2 demonstrates further examples of resistance measurements across the interferometer near 5/2 filling; alternation between e/4 and e/2 oscillations is evident in these results. Data from two different sample preparations are displayed, including resistance $\Delta R_L$ data, FFTs of these resistance measurements, and additionally plots of the differences in peak to peak positions versus the change in side gate voltage. In all $\Delta R_L$ data panels the e/4 and e/2 periods as marked by vertical lines are derived from FFT spectra peaks of the A-B measurements, and are consistent with A-B oscillation periods for their respective sample preparations at $\nu = 2$ and $\nu = 7/3$ or 5/3.

Panel (i) of Figure 2, preparation c, shows resistance $\Delta R_L$ versus side gate voltage and demonstrates distinct e/4 oscillations flanked by smaller amplitude e/2 oscillations. The data span 140 mV in side gate sweep and are divided into three sections each with a predominant oscillation period of e/2 or e/4. Each section, and the entire range, have FFTs applied with the results in panel (ii). Here the FFT spectrum of the full 140mV side gate sweep shows two principal peaks corresponding to e/4 and e/2 periods. The FFT spectrum of section I shows predominance of the higher frequency peak corresponding to the e/2 period oscillations. Section II FFT spectrum shows predominance of the e/4 period. Section III FFT shows e/2 period almost exclusively. These FFTs of the three separate sections demonstrate that the predominant period alternates between e/4 and e/2 over this side gate range. To further examine this alternation panel (iii) shows a plot of the difference in peak positions of adjacent prominent peaks throughout the side gate sweep extracted from the $\Delta R_L$ plot of panel (i). These data are the difference in peak position values of the series of oscillations shown in panel (i); [Vs of peak (n+1) - Vs of peak (n)] versus Vs of peak (n). This data demonstrate the relatively abrupt switching between the two periods of e/4 and e/2.

Sample preparation b, Figure 2, shows another example of this alternation between e/4 and e/2 periods, but with more alternations present. In this side gate range of roughly 180mV, the predominant period changes four times. Further examples of the e/4 and e/2 alternation are shown in the Supplementary Information Figure S3.

The data of Figure 2 are representative of the A-B oscillations present near 5/2. For all side gate sweeps of this range examined (Vs excursion of roughly 150 mV or more) and at lowest temperatures (~25mK), e/4 and e/2 oscillations are present and an alternation between the periods occurs. Measurements show that for different sample preparations the pattern of alternation and the Vs extent for each e/2 or e/4 sequence is different. The e/4 to e/2 alternation is aperiodic in that the extent of the e/4 or e/2 oscillations is not a fixed range, either within a sample preparation or when comparing two different sample preparations. To further examine this aperiodicity, measurements employing a larger range of $\Delta Vs$ are used.

Data from measurements applying a large range side gate sweep up to 350 mV are shown in Figure 3, again presenting the principal finding of alternation between e/4 and e/2 oscillations. Shown for are the plot of $\Delta R_L$ versus change in side gate voltage, plots of the adjacent peak position differences, and additionally the results of scanned FFTs over the full excursion range of the side gate sweep. The scanned FFT data is the result of performing an FFT over a window of ~30 mV progressing across the side gate sweep range at ~5mV steps (see Supplementary Information). The data plotted are the values from the FFT spectra at the frequencies corresponding to e/4 and e/2 A-B periods.

For the sample preparation shown in Figure 3, using a side gate sweep of more than 300 mV, the $\Delta R_L$ trace shows five sections of alternating e/4 and e/2 periods. Assignment of e/2 or e/4 period is supported by the adjacent peak position plots and by the scanned FFT plots, both demonstrating the alternation in periods. In this data with the larger total Vs excursion, the e/2 or e/4 oscillation range (e.g. e/4, 40 to 130mV in this sample preparation) can extend up to and more than 100mV, but neither the e/2 or e/4 sections are of fixed extent. From these large side gate sweeps the A-B interference can be summarized as an alternation of e/2 and e/4 periods occurring aperiodically. Data from two additional sample preparations are presented in Supplementary Information Figure S4.

A crucial property of these e/4 and e/2 period patterns is that they are reproducible. A large Vs sweep as in Figure 3 is not reproducible in the detail of the phase of the oscillations, but importantly the presence of e/4 or e/2 sections at a given specific side gate value is reproducible. For a given sample preparation the presence of e/2 or e/4 oscillation periods is dependent upon the value of the voltage applied, Vs: repetition of side gate scan over the same side gate voltage values will result in the same alternation pattern with extent of the e/4 or e/2 sets similar. This indicates that the presence of e/4 or e/2 is dependent upon the specific area enclosed by the interferometer. An example of this data reproducibility is shown in Supplementary Information Figure S5.

These data of alternating e/4 and e/2 period oscillations as a function of swept side gate voltage Vs are consistent with the proposed picture of 5/2 excitations possessing non-Abelian statistics. In this theory [16-20] for the excitations of e/4 charge traversing an interferometer as used here, a period corresponding to the e/4 charge would be expressed for a side gate voltage sweep where the number of encircled, localized quasiparticles is even and does not change during the sweep. If the encircled number of localized quasiparticles is odd, the e/4 oscillations are suppressed due to the non-Abelian nature of the quasiparticles. The data are consistent with this even-odd model. As the side gate is swept over a large range the number of localized quasiparticles in the enclosed area will change. The e/4 A-B oscillations shown in the data are where an even number of localized quasiparticles is present in the enclosed area. The e/2 oscillations experimentally displayed here in alternation with the e/4 oscillations represent the encircling of an odd number of quasiparticles within the interferometer area A. The possible mechanisms for e/2 oscillations are described below. More importantly, the alternation presents in the form expected for excursion of the side gate producing alternating even and odd total numbers of enclosed localized quasiparticles. The variation in the range of the gate voltage between switching or alternation back and forth from e/4 to e/2 is aperiodic or random in side gate voltage extent, and corresponds to the spatial distribution of localized e/4 quasiparticles, which should be random: The pattern of e/4 and e/2 oscillations provide a "finger-print" of the quasiparticle localization in the device. It is observed that the patterns of gate sweep extents of e/4 and e/2 oscillations are reproducible within a sample preparation (Supplementary Information Figure S5). This indicates that the localization potential landscape does not change for a given sample preparation, roughly reproducing the sites at which the quasiparticles are localized even as the gate is swept past depletion and repopulation at the localization site.

We turn now to altering the enclosed quasiparticle number by changing the applied B-field. This manipulation of the enclosed localized quasiparticle number between even and odd induces interchange between e/4 and e/2 A-B oscillation periods.

In these correlated electron systems, changing the applied B-field induces a change in the number of quasiparticles in the system, and near a quantum Hall state changes the number of localized quasiparticles. Changing the magnetic field over a sufficiently large range will therefore alter the quasiparticle population within a defined area, altering the parity of the number of quasiparticles between even and odd within that area. As described above, for a particular fixed B-field value, a side gate sweep shows the aperiodic alternation of e/4 and e/2 A-B oscillations that reflects the localized quasiparticle population within the area change of the interferometer: the alternation pattern is specific to the particular range of the side gate sweep. In this purported non-Abelian quasiparticle system, adjusting the applied magnetic field is expected to result in a distinctive change to this alternation pattern [16-20]: a parity change in the encircled localized quasiparticle number should interchange the e/4 and e/2 periods. A small B-field adjustment insufficient to add a localized quasiparticle will not change the pattern of aperiodic e/4 and e/2 observed over the same side gate sweep. However, a B-field change of sufficient magnitude to produce an odd number change in the localized quasiparticle population of the enclosed area interferometer will have a specific effect; the e/4 and e/2 periods in the alternation patterns should interchange for a side gate sweep over the same interferometer area.

This interchange of e/4 and e/2 periods upon changing B-field is shown in Figure 4. The top two panels show standard $\Delta R_L$ measurement versus side gate voltage over the same values of Vs, but at two different B-field values, and the bottom two panels are the corresponding adjacent peak position differences. The two B-value data sets show results consistent with the A-B e/4 and e/2 period alternation demonstrated thus far in this study; four alternations produce patterns of five e/4 or e/2 sections. However, the B-field change of 19G used here induces interchange of the e/4 and e/2 periods. Distinct interchange of e/4 and e/2 periods occurs for all five sections of these A-B oscillations shown in this data with this change in B-field. The data support the model that the alternation finger-print seen with the side gate sweep can be precisely interchanged in periods by addition of B-field.

Further demonstration of period interchange and experimental determination of the B-field change corresponding to addition of a single localized quasiparticle is shown in Figure 5. Mapped here are $\Delta R_L$ vs. $\Delta Vs$ over the same Vs range for multiple small increments in magnetic field values showing interchange in e/4 and e/2 at a measurable B-field interval. The value of this interval is checked for agreement with known interferometer parameters at 5/2 filling, providing a quantitative point of consistency with the model of non-Abelian e/4 quasipariticles.

The magnetic field change necessary to induce interchange in periods for a specific sample preparation is shown in Figure 5. This sample preparation is different than that of Figure 4. In the Figure 5 data set, a simpler pattern of e/2 and e/4 alternation (bottom panel (a), e/2 to e/4 to e/2) is displayed over the side gate range of nearly 400mV. The values of the side gate excursion are duplicated for the nine B-field values.

Here, interchange of the e/2 and e/4 periods is shown to occur for the first change in B-field value (increase from B=65.420). The new e/2 and e/4 period pattern is maintained upon further increase in B-field for the next two values. Reversion to the pattern of the initial B-field (panel a) occurs at higher B (above B=65.436, panel d). This pattern holds for the next two higher B values, although a small departure from this systematic interchange is noted at B near 65.450G in the large side-gate voltage range; the source of this anomaly is not presently known. As B-field is again increased reversion to the prior pattern is induced (increase above B=65.459, panel g). This pattern is sustained for the next several B-fields applied. These data show that the e/2 and e/4 alternation patterns can be interchanged with B-field change; the B-field change necessary to induce the interchange can be extracted from this data. The size of the shaded regions in Figure 5 roughly delineates similar intervals of e/4 or e/2 sections, and corresponds to B~20G. This is the interval of B-field change necessary to add a single localized quasiparticle.

This interchange of the alternation patterns induced here by a magnetic field change of 20 Gauss = $\Delta B$ is quantitatively consistent with that expected for introducing a single localized quasiparticle to this interferometer at filling factor 5/2: The expected B-field increment necessary to add that quasiparticle to the confinement area A, and so change the parity of the encircled quasiparticle population, is derived as follows: Near filling factor 5/2 the ratio of electrons to flux quanta is $5/2 = e/\phi_o$, and the expected quasiparticle charge is $e^* = e/4$, or $e = 4e^*$, so that each $\phi_o$ corresponds to $10e^*$: $(1\phi_o)(5e/2\phi_o)(4e^*/e)$, or $0.1\phi_o/e^*$. For the interferometer area A~$0.2\mu m^2$ (derived in ref. 21 and consistent with the gate voltages used here), and magnetic flux quantum $\phi_o = 40$G-$\mu m^2$, the B-field change per quasiparticle is $\Delta\phi_o/e^* = (0.1\phi_o/e^*)(40$G-$\mu m^2/0.2\mu m^2) \sim 20$G/$e^*$. This number is consistent with the data of Figure 5 showing multiple e/4 and e/2 interchanges.

The finding of alternation of e/4 and e/2 oscillations as encircled area A is changed and the finding of interchange of e/4 and e/2 oscillation interference patterns with magnetic field change are as theoretically prescribed for non-Abelian e/4 quasiparticles. These interferometric results in tandem are supportive of the conclusion that e/4 quasiparticles are non-Abelian.

While these results provide consistency with the model of non-Abelian e/4 charges, the origin of the e/2 period is open. Two leading mechanisms for e/2 period oscillations are that Abelian e/2 charges are present and demonstrate interference, or that e/4 charge traverses the interferometer twice. Both pictures consider e/4 oscillations as A-B oscillations for an even enclosed quasiparticle number, and the e/2 oscillations are exposed due to <u>suppression</u> of e/4 oscillations. Interference by e/2 charge quasiparticles may be the preferred model: e/2 periods result from the presence of Abelian e/2 charge quasiparticles, exposing a fundamental charge in the system [26-27]. Several recent theoretical studies [27-29] have discussed the interference processes in these devices and explicitly examined the mechanisms of the e/2 oscillations. In the picture of e/2 period due to twice traversal of the area by e/4 charge, that quasiparticle must complete two laps around the interferometer, encircling an even quasiparticle number: the total area encircled is doubled, resulting in an e/2 period. This mechanism is unlikely given the additional tunneling events at the qpcs, markedly reducing the e/2 oscillation amplitudes.

To conclude, the data presented here show alternating e/4 and e/2 period oscillations that are distinctly consistent with the theory of non-Abelian e/4 quasiparticles. The mechanisms of changing area through side gate voltage or changing B-field produce this alternation and interchange. The alternation reflects the specific even-odd statistical property proposed for non-Abelian e/4 quasiparticles. The properties of these oscillations are as expected for the operation of the interferometer where a random spatial distribution of localized quasiparticles is traversed by the side gate. Addition of magnetic flux corresponding to addition of an odd number of localized quasiparticles is consistent with the observed interchange of e/4 and e/2 period oscillation sequences. The conclusion that e/4 quasiparticles are non-Abelian has important ramifications for the utility of this system; e/4 excitations are consequently viable candidates for performing topological quantum computation operations.

FIGURE CAPTIONS

FIGURE 1. <u>Interference device and Aharonov-Bohm oscillations at multiple filling factors, showing e/4 and e/2 periods near 5/2.</u> Panel (a): confined area $A_L$ is defined by top gates q, quantum point contacts, and the central channel gate c, with voltage Vs. Change in Vs independently controls the interferometer area A. $R_L$ measurement uses drive current from contacts 1 to 2 and voltage drop from 3 to 4. The inset to (b) is the device electron micrograph. Panels (c-e): $\Delta R_L$ versus side gate sweep $\Delta$Vs, showing Aharonov-Bohm (A-B) oscillations at filling factors ν = 2, 7/3 and 5/2. Vertical lines in the ν =2 and ν = 7/3 traces are consistent with the peaks of their respective fast-Fourier transforms (FFT) of the spectra (c and d insets). In A-B oscillations the period is dependent upon quasiparticle charge e* and magnetic field B with period~$\Delta$Vs~$\Delta$A~(h/e*)/B. In panel (b) measured period x B is plotted against the assumed charge for ν = 2 and 7/3 of e and e/3 respectively; the data are consistent with A-B oscillations.

      Panel (e): $\Delta R_L$ versus side gate sweep $\Delta$Vs for the same sample preparation at 5/2 shows two periods corresponding to quasiparticle charge e/4 and e/2 periods. Vertical lines marking these periods are consistent with FFT peaks taken over the range at which each predominates (panels f and g). Charge periods e/4 and e/2 are consistent with A-B oscillations as shown by the inclusion of their measured period and B products in Figure 1b: the measured periods agree with the expected values for charges e/2 and e/4.

      Data taken at T=25mK, and drive current 2nA, sample preparation a. All swept gate data displayed in this study are measured on the high B-field side of the indicated filling factor.

FIGURE 2. <u>Alternation between e/4 and e/2 period Aharonov-Bohm oscillations near 5/2 for two sample preparations.</u> Sample preparation b, top, panel (i): near 5/2, $\Delta R_L$ vs. $\Delta$Vs for total $\Delta$Vs~200mV; vertical line periods are from FFT peaks and correspond to e/4 and e/2 periods as determined by period measurements at filling factors 2, 5/3 or 7/3 for each preparation. Within these traces each section showing either e/2 or e/4 period is numbered and has a respective FFT of that section shown in panel (ii). Panel (ii): respective FFTs of panel (i) sections demonstrating predominance of either e/4 or e/2 periods in each section. Panel (iii): difference in adjacent predominant peak positions; peak position in side gate voltage ($V_i$) is subtracted from the next higher Vs peak position ($V_{i+1}$), plotting $V_{i+1}$-$V_i$ vs. $V_i$. Alternation of e/2 and e/4 oscillation periods are demonstrated in the two predominant adjacent peak position differences. The two sample preparations show different alternation patterns, with preparation b displaying five sections and preparation c three sections of e/4 and e/2 predominance. This alternation can be attributed to change in the parity of the number of quasiparticles enclosed in interferometer area A as Vs is swept to change this area, and the different alternation patterns reflect the two distinct spatial patterns of localized quasiparticles in each preparation. T=25mK and current=2nA in both preparations.

FIGURE 3. <u>Large range side gate sweep induced alternation of e/4 and e/2 oscillation periods.</u> Displayed are interferometer longitudinal resistance $\Delta R_L$ versus change in side gate voltage, swept window FFTs of the $\Delta R_L$ spectra, and adjacent peak positions. A large side gate voltage sweep of near 400mV displays four alternations between e/2 and e/4 period: the top trace is the $\Delta R_L$ measurement, with vertical line separation there corresponding to e/2 and e/4 periods as defined by period measurements at filling factors 2, 5/3 or 7/3 for each preparation. The middle trace is the FFT amplitude value corresponding to the peak in the FFT spectrum of either e/2 (black) or e/4 (red) swept over the full side gate range. The FFT uses a window of roughly 30mV and progresses at 5mV steps over the spectrum of the top trace. The FFT values as a function of side gate voltage show alternation of e/4 and e/2 periods as the side gate is swept. The bottom trace is the difference in adjacent peak positions for the predominant peaks in each section of either e/2 or e/4 periods.

The aperiodic e/4 and e/2 alternation can be ascribed to sweeping over a population of localized e/4 quasiparticles, alternating an enclosed even-odd number, and expressing the e/4 oscillation periods according to the quasiparticle's non-Abelian properties. The Figure's schematic depicts this process.

T=25mK, current 2nA, preparation d. Additional data from other sample preparations are displayed in the Supplementary Information Figure S4.

FIGURE 4. <u>B-field induced interchange of e/4 and e/2 oscillation sequences near filling factor 5/2.</u> Top two panels are $\Delta R_L$ versus change in side gate voltage Vs showing A-B oscillations corresponding to charge e/4 and e/2 periods. The two traces extend over the same side gate voltage range of -4.00V to -4.35V, but are at different magnetic field values of B = 65.500kG and 65.519kG. With addition of this small B-field the pattern in the top trace of e/2 and e/4 periods is replaced with a pattern in which e/2 and e/4 are interchanged. This interchange of e/2 and e/4 patterns is further displayed in the bottom two panels where the adjacent peak differences of the top traces are plotted; the interchanges of e/2 and e/4 occur for all five period sections present in these traces.

This change of 19G can correspond to addition of one or an odd number of quasiparticles to the area of the interferometer, changing the parity of the number of enclosed quasiparticles. For the e/4 quasiparticle to be non-Abelian such a change in parity would induce this observed pattern of interchange of the e/2 and e/4 periods. T=25mK and current = 1nA, sample preparation e.

FIGURE 5. <u>Interchange of e/4 and e/2 period Aharonov-Bohm oscillations near 5/2 showing the B-field increment that adds one localized e/4 quasiparticle to the interferometer.</u> These peak to peak plots are derived from $\Delta R_L$ measurements in side gate sweeps of ~ -4.5V to -4.8V at nine different B-field values. The B-field values progress in ~ 8G increments, demonstrating four interchanges of the e/4 and e/2 periods. The shaded (non-shaded) regions correspond to the presence of predominant e/2 (e/4) periods. The period of interchange is ~ 20 G, corresponding to the B-field increment necessary to add a single localized e/4 quasiparticle to the confinement area, changing the parity of the encircled quasiparticle population. This interchange and the 20G B-field increment sufficient to induce it are consistent with expected non-Abelian properties of e/4 quasiparticles and with the particular parameters of this interferometer at 5/2; see text. The data are all taken with current = 1nA and T= 25 mK, sample preparation f.

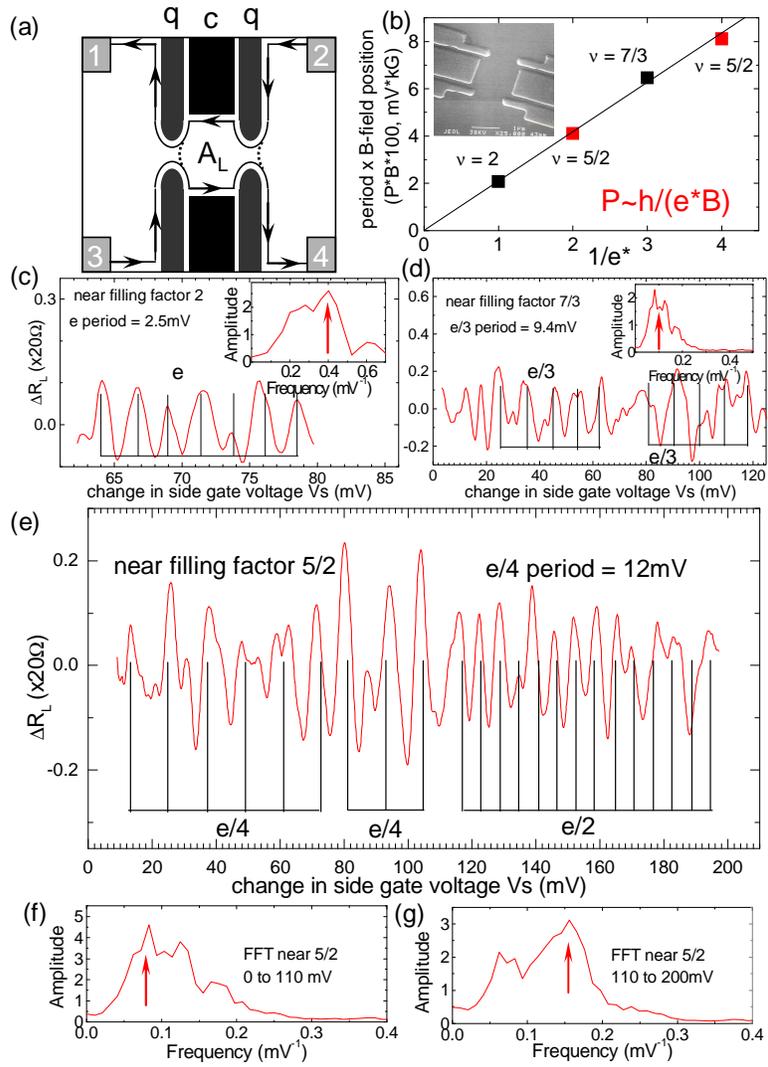

Figure 1.

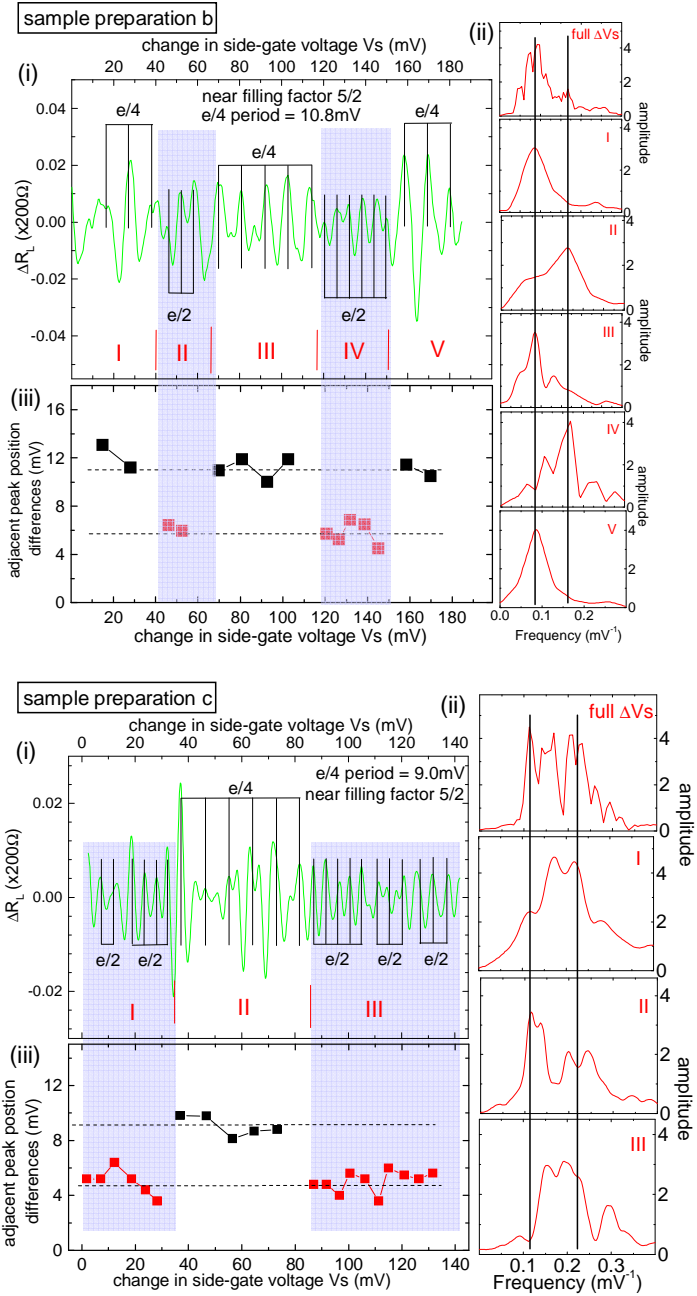

Figure 2.

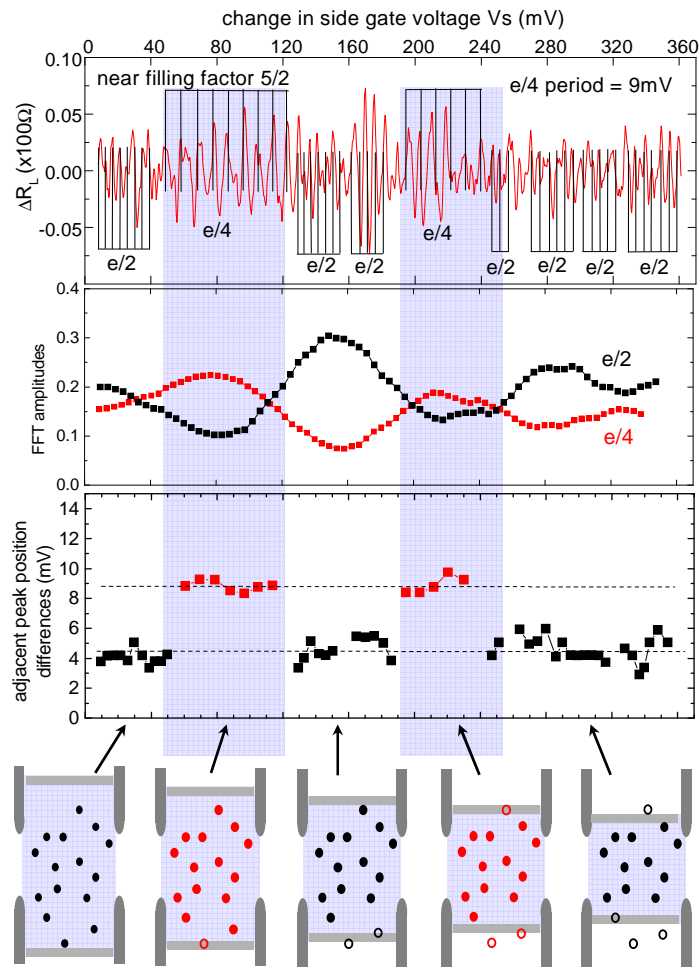

Figure 3.

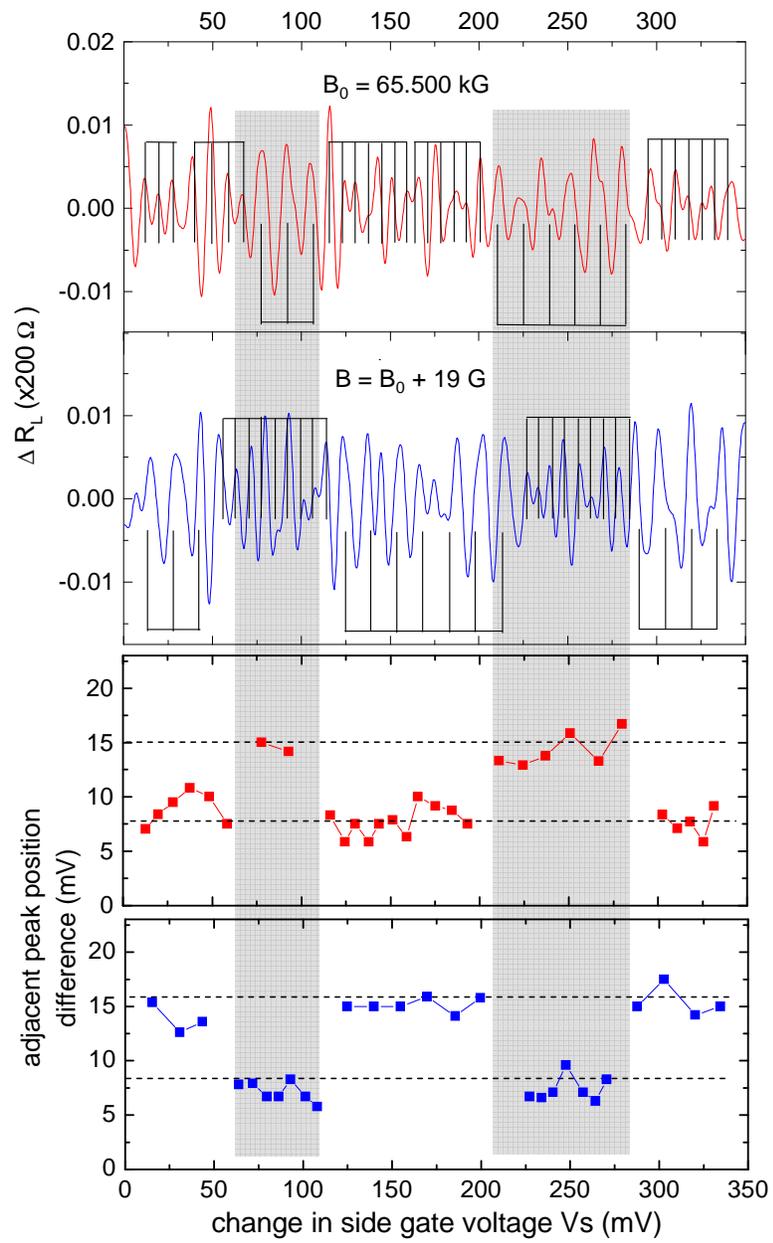

Figure 4.

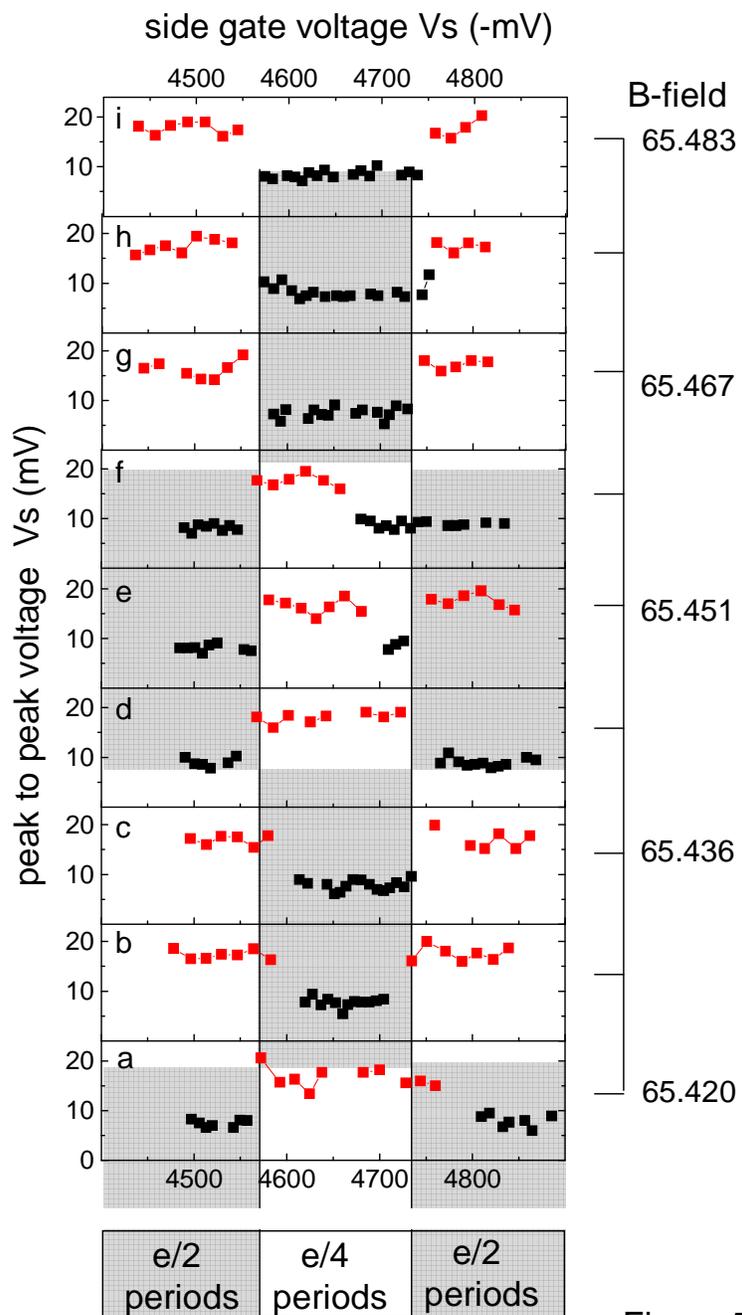

Figure 5.

Supplementary Information for

# Alternation and interchange of e/4 and e/2 period interference oscillations as evidence for filling factor 5/2 non-Abelian quasiparticles

R.L. Willett*, L.N. Pfeiffer, K.W. West
Bell Laboratories, Alcatel-Lucent
600 Mountain Avenue
Murray Hill, New Jersey, 07974 USA
Correspondence to: rlw@alcatel-lucent.com

Supplementary Information:

Methods:

High mobility, high density heterostructures with Ni/Au/Ge contacts diffused into the mesa edge are used here as in previous measurements [21, 24]. The heterostructures are of fixed density n away from the top gate structures, with filling factors $\nu = nh/eB$ adjustable through sweep of the magnetic field B. The interferometer is a top gate structure with biasing able to deplete the underlying 2D electron system and adjust the 2D electron density laterally adjacent to the gates. It is shown schematically in Figure 1a, and is composed of two quantum point contacts (qpcs) and a central channel region. Each component is independently controllable, with previous studies [21, 24] detailing the operation of the device and its components. The 2D electron gas is approximately 200nm below the heterostructure surface. A 40nm SiN layer on the surface of the heterostructure further separates the 2D electron gas from the Al top gates. These gates fully deplete the underlying 2D electron gas at about -2.5V; this top gate structure is displayed in the electron-micrograph in Figure 1b inset. Transport is measured through the devices using standard lock-in techniques (see below), with $R_L$ [25] as defined in Figure 1. The samples are illuminated with a red LED at low temperatures, down to 25mK in a dilution refrigerator.

Data displayed in this study include results from a total of eight sample cool-downs (room temperature to 25mK) with multiple illuminations within each cool-down. The red LED illumination empirically serves to enhance the sample mobility and induces differences in the sample conduction, even in bulk transport examinations. These illuminations therefore produce a nominally different sample preparation with each illumination cycle.

The experiments are conducted in a window of qpc operation of sufficiently low bias that allows the FQHE states to exist, but large enough bias to observe interference. This principal is discussed below.

A fundamental aspect of these experiments is that the interferometer is to be operated such that the fragile fractional quantum Hall states at 5/2, 7/3 etc. are preserved within the device, yet sufficient backscattering must be induced to see quasiparticle interference. The backscattering is produced at the qpcs, where ideally a path of density uniform with the bulk is maintained and backscattering occurs across this density. Inducing this backscattering for integral quantum Hall states where the energy gaps are >10K is readily achievable given those states are robust against density perturbations: the edges can be brought into close proximity to achieve high backscatter rates across a small tunneling dimension. This is not the case for the fragile FQHE states with gaps two orders of magnitude smaller: the qpcs must be kept open to sustain the bulk density in the gap (while still fully depleting beneath the gate), and the tunneling must therefore occur over a larger dimension. This larger tunneling distance is the root of the small amplitude, phase interrupted interference pattern that is observed in this study. The similar interference noted for the integers as for the fractions reflects the fact that the tunneling is occurring over the same gate biases: in all measurements here the qpcs are maintained at the same biases when comparing integer and fractional states in order to retain the dimension A of the interferometer. Large biases on the qpcs would induce better integral state interference but damage the FQHE states.

The sample preparation involves a procedure of charging the gates, then allowing the system to equilibrate. The oscillatory features at FQHE states at 5/3, 7/3 and at 5/2 filling factor are improved with a period of equilibration of several days. This procedure, in addition to noise reduction efforts, have allowed closer examination of the interference of FQHE states compared to previous work [21]. Experimental modifications and procedures were also undertaken from previous experiments to allow large side gate excursion.

Small signal sensitivity is achieved by very slow sweep of the side gate ($\Delta V_s$) at rates typically of ~400mV/24 hours using long time constants (30 to 100 seconds) in the lock-ins and the data are averaged as taken over 0.2mV wide bins; different traces cannot be averaged due to phase changes between sweeps, however the collected data is averaged instrumentally using this standard technique.

Data analysis detail summary:

Extraction of oscillatory properties of $R_L$ data included fast Fourier transforms of the $\Delta R_L$ data and determining adjacent peak position differences, with marking of periods on the $\Delta R_L$ data utilizing those methods for extracting periods.

The vertical line period markings placed on the $\Delta R_L$ data are adjusted in phase to accommodate the series of oscillations present. As noted in previous studies [21], oscillation series typically do not present as long continuums of oscillations, but instead are usually series of several to many oscillations separated by phase disruptions. The markings placed on the $\Delta R_L$ data use the periods derived from the FFTs and from the control measurements at 2, 5/3 and 7/3, but are placed on the data consistent with these phase disruptions. To indicate a continuous series of oscillations, the vertical markings for such a set are linked by a horizontal line and labeled with the appropriate period charge.

Fast Fourier transforms (FFTs) performed on the $\Delta R_L$ data are the principal means of extracting the predominant oscillation periods. While used on all filling factor data, for the smaller quasiparticle charge oscillations (larger periods) the minimum range of Vs sweep over which periods could be well determined is limited. It is observed that less than 15mV Vs range is not viable for extracting oscillation frequencies, and that at 25mV periods were defined and resolved. This finding sets the limit on swept FFTs (Figure 3), and the larger frequency resolution for large Vs extent prompted examining sections of the full Vs scans showing predominant periods on inspection (Figure 2).

The second analysis method for examining the oscillatory properties in the data uses the differences in adjacent peak positions.  Here the peak position in side gate voltage of a given peak ($V_i$) is subtracted from the next higher Vs peak position ($V_{i+1}$), with these values plotted over the full range of the side gate sweep ($V_{i+1}-V_i$ vs. $V_i$).   The predominant peaks within sections (e/4 or e/2 prevalent periods) of Vs sweep are used, neglecting lower amplitude features.  Examples of this are described below.

FFTs are performed on either the full $\Delta R_L$ range displayed in each figure or on the range explicitly delineated with each spectrum.  The FFT has the advantage that the full spectral weight of $\Delta R_L(Vs)$ is used in each transform. The principal disadvantage to the FFTs is the finite extent of Vs needed to derive a meaningful FFT spectrum, so that abrupt changes in oscillatory period are not readily apparent.  For assessing prevalent periods the adjacent peak difference method overcomes this limitation in the FFT, but does not incorporate the full spectral weight of the data: the peak difference period appears intrinsically noisier than the period definition by FFTs.

Data analysis methods applied to each figure are outlined below.

Figure 1: Plots of $\Delta R_L$ take the measured $R_L$ and subtract the gross background values at that filling factor. Fast Fourier transforms (FFTs) in the insets of (c) and (d) are taken over the entire range of data shown in the respective $\Delta R_L$ traces.  The FFTs covering the $\Delta R_L$ measurement of (e) are taken over the ranges listed on the figure.

Figure 2: FFTs in the (ii) panels are extracted from the sections defined in the (i) panels.  The adjacent peak positions plot data is derived by subtracting the Vs position of a peak from the Vs position of the next peak at higher Vs.  The threshold for defining a peak is larger in the e/4 sections than in the e/2 sections.  An example is Figure 2, preparation b, at ~85mV.  This peak is not included in the peak differences of panel (iii).  For the two data sets of Figure 2 and Supplementary Information Figure S3 roughly four such additional features are present within a total of nearly eighty peaks.

Figure 3: The scanned FFT employed here is comprised of FFTs taken of a window roughly 30mV wide, stepped across the $\Delta R_L$ trace at 5mV increments.  The data plotted are the FFT amplitude at the e/2 peak frequency position (black) and at the e/4 peak frequency position (red) where these positions are determined by the peaks apparent in FFT of the entire $\Delta R_L$ spectrum trace.  To properly normalize the e/2 and e/4 amplitudes for the different FFTs, four FFT amplitudes are taken at four different frequencies (mV$^{-1}$), and their sum is used to normalize the e/4 and e/2 points.  These normalized values are plotted in the (ii) panels after adjacent amplitudes are averaged in each.

Figures 4 and 5:  all analysis here is described previously.

Supplementary Information Figure Captions:

Supplementary Information Figure S1.
Four terminal resistance near and through the interferometric device.  Top panel is longitudinal resistance measured adjacent to the interferometric device in the bulk of the 2D electron system. Bottom panel is $R_L$ measured across the device as described in Figure 1 of the main text. T=25mK, current 2nA, sample preparation g.

Supplementray Information Figure S2.
<u>Aharonov-Bohm oscillation periods for magnetic field values near integer filling factors of 2, 3, 4, 6, and 8</u>. The periods increase inversely as 1/B, as expected for A-B oscillations in which oscillation period ~ (h/e*)/B. Sample $\Delta R_L$ measurements are displayed in the right side panels, noting that the Vs axis is changing by a multiple of 2 in the progression. These results are inconsistent with Coulomb blockade but are as expected for A- B oscillations; see results of ref. 26. Additionally, measurements of oscillations at different integers for swept B-field show no change in period over this filling factor range, also consistent with A-B oscillations. The geometry of the devices used here is likely responsible for the presence of A-B oscillations: the open qpcs promote A-B oscillations rather than Coulomb blockade. T=25mK, I = 1nA, sample preparation h.

Supplementary Information Figure S3.
<u>Alternation of e/4 and e/2 period oscillations over side gate sweep of roughly 160mV</u>, with FFT resolution of $\Delta R_L$ measurement sections. Panel (i) shows $\Delta R_L$ measurement versus side gate voltage change over 160mV. The marked vertical lines are consistent with the period peaks determined by the FFTs shown in (ii) where the sections marked in (i) have been separated and Fourier transformed. These marked vertical lines (periods) are corroborated as A-B oscillations in similar measurements at filling factors 2 and 7/3, where the periods marked here are consistent with charges of e/2 and e/4 as determined by the periods at 2 and 7/3. The predominant peak to peak position differences in the sections are displayed in (iii), demonstrating the prevalence of the two periods corresponding to e/4 and e/2 charges and their alternation.

    Measurements were performed at temperature = 25mK and with 2nA current; sample preparation i.

Supplementary Information Figure S4.
<u>Alternation of e/4 and e/2 periods over a large side gate sweep range</u>, demonstrating resolution of e/4 and e/2 period sections in two sample preparations. For each sample, the $\Delta R_L$ measurement is shown in the top trace, scanned FFT over this spectrum is shown in the middle panel, and adjacent peak to peak position differences are shown in the bottom panel for a single sample preparation. As in the main text Figure 3, the scanned FFT uses a window of approximately 30mV, with FFT performed at ~5mV increments. The red and black data of the middle panel correspond to FFT amplitude peak values at e/4 and e/2 respectively for the window. The normalized e/4 and e/2 period FFT amplitudes alternate in which is larger over the range of side gate sweep, corresponding to the e/4 and e/2 alternation in the data of the bottom and top panels.

    Measurements taken at temperature = 25mK, 2nA current; sample preparations k and l.

Supplementary Information Figure S5.

Reproducibility of e/4 to e/2 sequences in longitudinal resistance $\Delta R_L$ measurements. Four side gate voltage sweeps traversing the same total applied side gate voltage at the same single fixed B-field near filling factor 5/2 are shown in the figure, each from -3.600V to -3.950V. Each trace A-D corresponds to slow application (~ -15mV/hr) of bias to the channel gate set Vs (Figure 1) over the range of the abscissa with similar return rate to the initial bias value. The e/2 and e/4 periods are marked for comparison between traces, with an error of starting bias of roughly +/- 5mV between traces. Note that the general pattern of the e/4 and e/2 sequences is reproducible and maintained in some detail between different traces; all four traces show the same oscillation periods of e/2 or e/4 for a given total side gate voltage within the error of the applied starting voltage. Four e/4 sequences occur from roughly 0 to 40mV, 100mV to 150mV, from near 175mV to 225mV, and again at 300mV. In the 100 to 150mV sequence the six peaks are generally in phase with respect to traces A to D. However, in the 175 to 225mV sequences six peaks again occur but are out of phase. Note the sections of apparent phase disruption (150 to 170mV, and 225 to 245mV) where assignment of period is not readily possible.

      Beyond the properties of the aperiodicity and reproducibility, the general finding [21] that the e/2 oscillations are of smaller amplitude than the e/4 is again supported. Another property of the e/2 sections apparent in these data is that *within* each section the runs or series of e/2 oscillations are generally of smaller duration than the e/4 runs within a section; the phase disruptions are more prevalent in the e/2 sections. These large Vs excursion data demonstrate that alternation of e/4 and e/2 period oscillations is a dominant feature of the 5/2 interferometric result, with particular properties exposed that include the aperiodicity and reproducibility of the alternation, and the lower amplitude and smaller oscillation sequences of the e/2 versus e/4 oscillations. These phase disruptions have been discussed in previous reports [21]. T=25mK, current 2nA, sample preparation j.

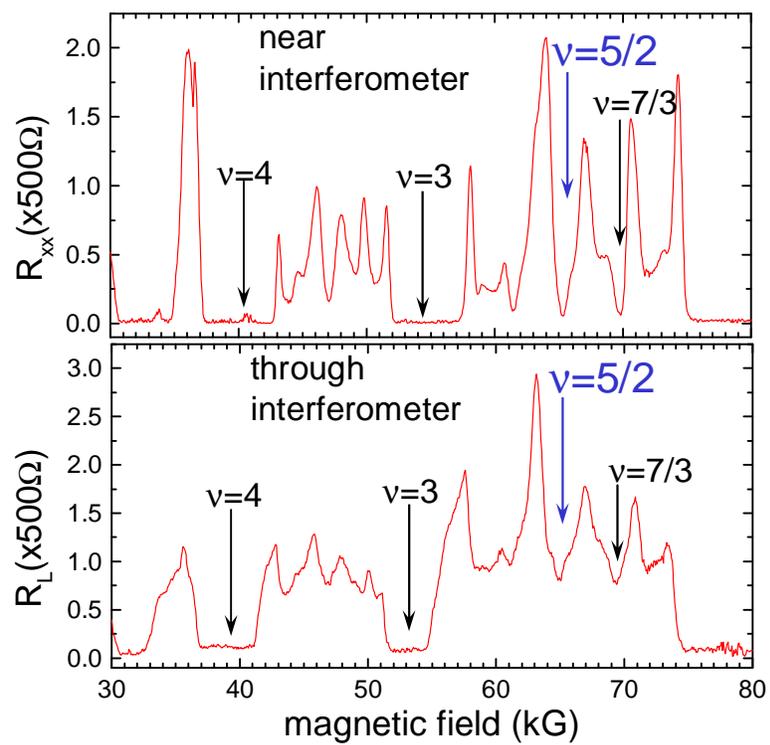

Supplementary Information Figure S1

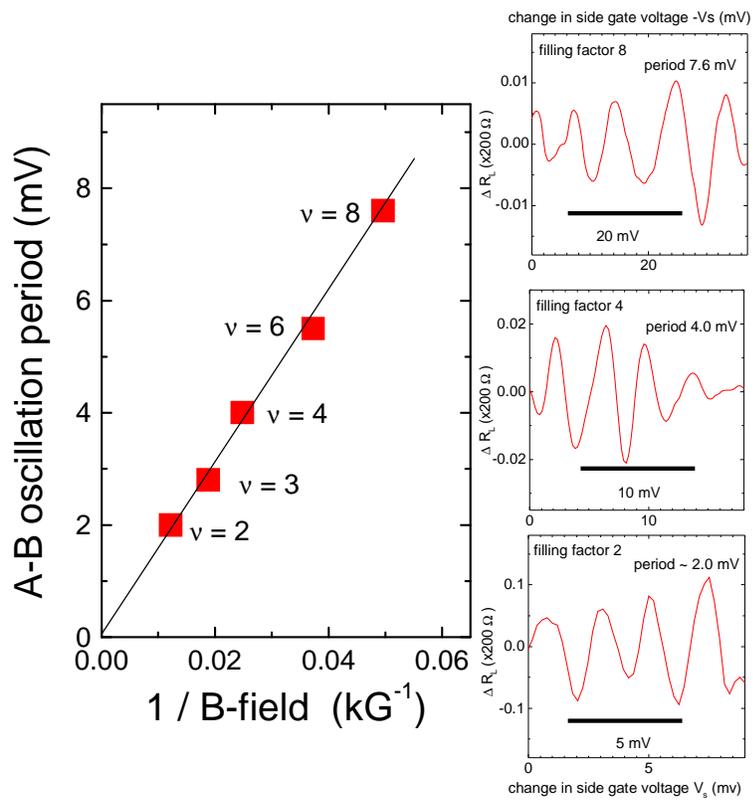

Supplementary Information Figure S2

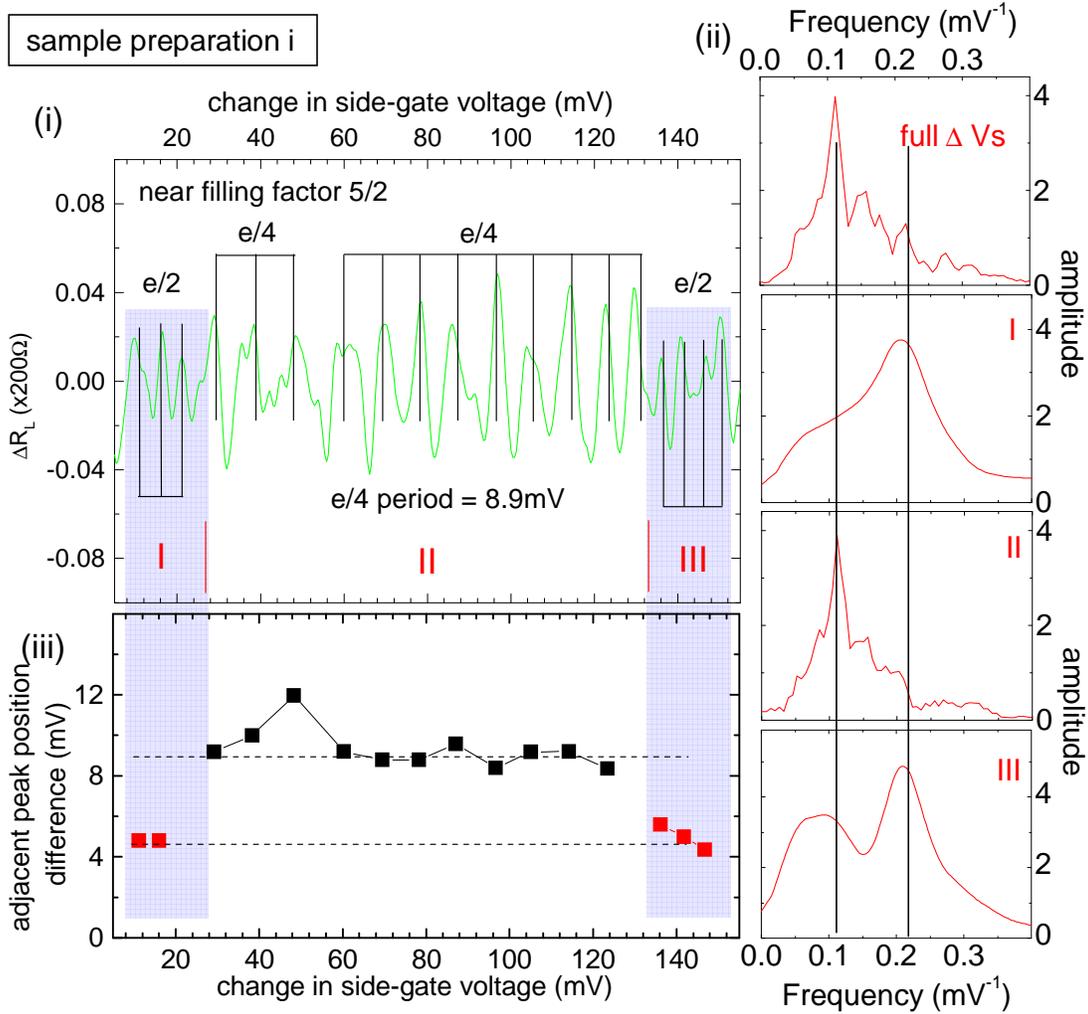

Supplementary Information Figure S3.

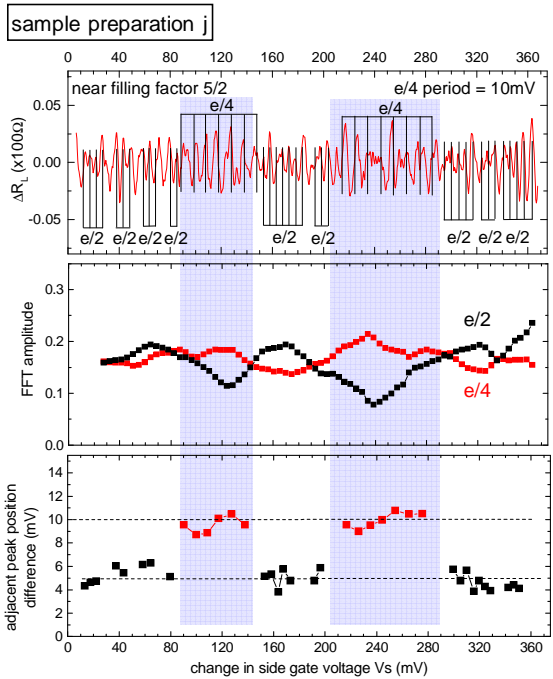

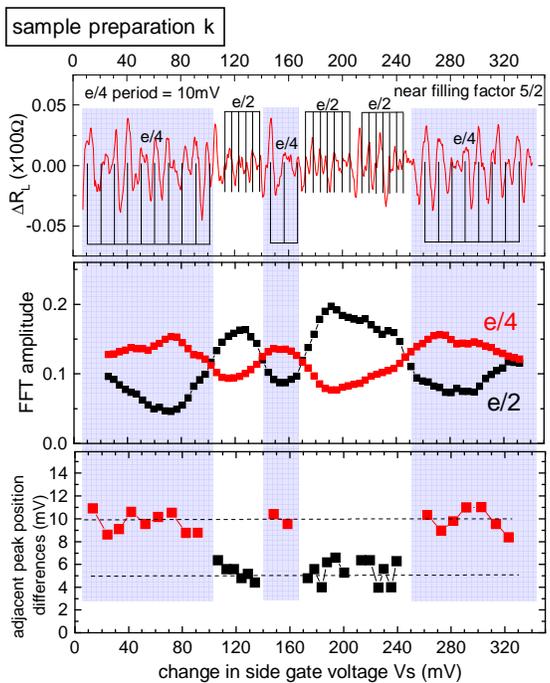

Supplementary Information Figure S4.

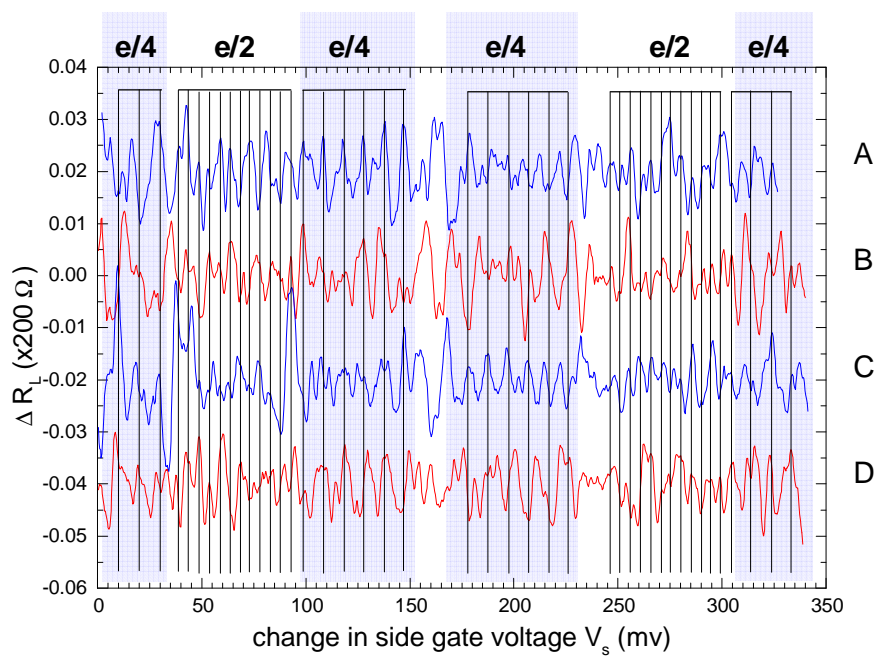

Supplementary Information Figure S5.